\documentstyle[preprint,aps]{revtex}

\begin{document}
\title{Unified order-disorder vortex phase transition in high-T$_{c}$
superconductors}
\author{Y. Radzyner, A. Shaulov, Y. Yeshurun}
\address{Department of Physics, Institute of Superconductivity, Bar-Ilan\\
University, Ramat-Gan, Israel}
\date{October 30, 2001}
\maketitle

\begin{abstract}
The diversity of vortex melting and solid-solid transition lines measured in
different high-T$_{c}$ superconductors is explained, postulating a unified
order-disorder phase transition driven by both thermally- and
disorder-induced fluctuations. The temperature dependence of the transition
line and the nature of the disordered phase (solid, liquid, or pinned
liquid) are determined by the relative contributions of these fluctuations
and by the pinning mechanism. By varying the pinning mechanism and the
pinning strength one obtains a spectrum of monotonic and non-monotonic
transition lines similar to those measured in Bi$_{2}$Sr$_{2}$CaCu$_{2}$O$%
_{8}$, YBa$_{2}$Cu$_{3}$O$_{7-\delta }$, Nd$_{1.85}$Ce$_{0.15}$CuO$%
_{4-\delta }$, Bi$_{1.6}$Pb$_{0.4}$Sr$_{2}$CaCu$_{2}$O$_{8+\delta }$ and (La$%
_{0.937}$Sr$_{0.063}$)$_{2}$CuO$_{4}$.
\end{abstract}


Vortex matter phase transitions have been under close scrutiny in recent
years. Both experimental\cite{1,2,3,4,5,6,7,8} and theoretical\cite{9,10,11}
works have indicated the existence of two order-disorder phase transitions:
A transition from a quasi-ordered solid phase to a liquid phase driven by
thermal fluctuations, and a transition to a disordered solid phase driven by
disorder-induced fluctuations. In magnetization experiments, the melting
transition is signified by a jump in the reversible magnetization\cite{1},
whereas the solid-solid transition is associated with the appearance of a
second magnetization peak\cite{2,6} (`fishtail'). A variety of experiments
indicate that the melting\cite{1} as well as the solid-solid transition \cite%
{12,13} are of first order.

While melting lines measured in different samples exhibit qualitatively
similar behavior, with the melting field decreasing monotonically as
temperature is increased\cite{1,3}, the solid-solid transition lines
measured in different samples differ markedly: A flat transition line in
underdoped Bi$_{2}$Sr$_{2}$CaCu$_{2}$O$_{8}$(BSSCO)\cite{14}, which
terminates at intermediate temperatures; a flat transition line followed by
a monotonic convex decrease toward T$_{c}$ in Bi$_{1.8}$Pb$_{0.8}$Sr$_{2}$%
CaCu$_{2}$O$_{8+\delta }$\cite{15} and Nd$_{1.85}$Ce$_{0.15}$CuO$_{4-\delta
} $ (NCCO)\cite{6}; a steep concave decrease throughout the whole
temperature range in (La$_{0.937}$Sr$_{0.063}$)$_{2}$CuO$_{4}$ (LaSCO)\cite%
{8} and some YBa$_{2}$Cu$_{3}$O$_{7-\delta }$(YBCO) samples\cite{3,16}; and
a non-monotonous behavior exhibiting a peak in YBCO\cite{3,4,5,13,17}, BSCCO %
\cite{14,18,19} and Bi$_{1.6}$Pb$_{0.4}$Sr$_{2}$CaCu$_{2}$O$_{8+\delta }$
(Pb-BSCCO)\cite{7}. The diverse temperature dependence of the vortex
solid-solid transition line is illustrated in Fig. 1 for YBCO\cite{13}, NCCO %
\cite{6} and LaSCO\cite{8} samples measured in our laboratory.

Both the melting and the solid-solid transitions may be observed in the same
sample in different temperature regimes. The melting line, appearing in the
high temperature region, terminates at a ``critical'' point\cite{2,20} and a
second line, associated with the solid-solid transition, emerges\cite{2,6}.
Recent experiments\cite{18,21} demonstrated that the two transition lines
are in fact a single line along which order is destroyed; the melting of the
quasi ordered solid into a liquid at high temperatures changes its character
into a solid-solid transition at low temperatures due to slower dynamics.
Striking evidence for the unified nature of these two lines was recently
found in vortex ``shaking'' experiments which show that by enhancing
relaxation effects, the second peak anomaly is transformed into a jump in
reversible magnetization\cite{18}, demonstrating that the melting and the
solid-solid transition lines are different manifestations of the {\it same}
phenomenon, i.e. a transition from an ordered phase to a disordered phase.

Motivated by the above results, and based on a recent theoretical model \cite%
{9,10,11}, we present in this paper a unified approach to the vortex
order-disorder phase transition, postulating that this transition is driven
by {\it both} thermal and disorder-induced fluctuations \cite%
{11,23,23p2,23p5,24}. Our simplified analysis is capable of reproducing the
markedly different behavior of the transition lines observed experimentally
in different samples. A spectrum of different transition lines, with
monotonic or non-monotonic behavior, is obtained by tuning the pinning
strength incorporated into different pinning mechanisms.

A recent model \cite{9,10,11} applies the Lindemann criterion to define a
transition from an ordered state to a disordered one. Previous approaches to
this model (e.g. \cite{4,5,6,10}) commonly dealt with the melting and the
solid-solid transitions separately, postulating that the former is driven by
thermal fluctuations and the latter by disorder-induced fluctuations.
Accordingly, the melting line $B_{m}(T)$ was determined \cite{9,10} by a
competition between the vortex lattice elastic energy and the thermal
energy, whereas the solid-solid transition line $B_{ss}(T)$ was determined
by a competition between the elastic energy and the pinning energy.
Following this approach one encounters several difficulties. For example,
one cannot explain the effect of point defects on the melting line observed
experimentally\cite{19,22}. In addition, this approach cannot explain the
wide spectrum of qualitatively different solid-solid transition lines
obtained in different materials, and even in different samples of the same
material. In particular, contrary to the predictions of the model, which
dictates a temperature independent $B_{ss}\left( {T}\right) $ at low
temperatures, a wide spectrum of temperature dependences is observed
experimentally\cite{3,4,5,6,7,8,13,14,15,16,17,18,19}.

The above difficulties are resolved by considering the effect of {\it both}
thermal fluctuations and disorder-induced fluctuations in destroying the
vortex lattice. The basic premise of our analysis is that an order-disorder
transition occurs when the sum of the average thermal and the
disorder-induced displacements of the flux line, $u_{T}^{2}$ and $%
u_{dis}^{2} $, respectively, exceeds a certain fraction of the vortex
lattice constant{\sf \ }$a_{o}$\cite{11,24}. A more accurate analysis should
involve the averaged total displacement of the flux line, which is not
necessarily the sum of $u_{T}^{2}$\ and $u_{dis}^{2}$. Yet, our simplified
approach yields a qualitative description, and provides important insight.
Utilizing the Lindemann criterion for the destruction of order, the
transition line, $B_{OD}(T)$, will obey the expression:

$u_{T}^{2}\left( {L_{o},0}\right) +u_{dis}^{2}\left( {L_{o},0}\right)
=c_{L}^{2}a_{o}^{2}$ \qquad \qquad (1) \newline
where $u_{T}^{2}\left( {L_{o},0}\right) =L_{o}kT/\left( 2\epsilon
_{o}\varepsilon ^{2}\right) $ is the transverse excursion caused by thermal
agitation and $u_{dis}^{2}\left( {L_{o},0}\right) =\left( \xi ^{2}/2\right)
\left( {L_{o}/L_{c}}\right) ^{6/5}$ is the disorder-induced fluctuation.
Here, $L_{o}=2\varepsilon a_{o}$ is the characteristic length for the
longitudinal fluctuations, $L_{c}=(\varepsilon ^{4}\epsilon _{o}^{2}\xi
^{2}/\gamma )^{1/3}$ is the size of the coherently pinned segment of the
vortex \cite{25}, $\epsilon _{o}=(\phi _{o}/4\pi \lambda )^{2}$ is the
vortex line tension, $\varepsilon =\sqrt{m_{a}/m_{c}}$ is the anisotropy
ratio, $c_{L}{}$ is the Lindemann number, $a_{o}=\sqrt{\phi _{o}/B}$ is the
Abrikosov lattice constant, $\phi _{o}\simeq 2.07\times 10^{-7}$ $G\cdot
cm^{2}$ is the flux quantum, and $\gamma $ is the pinning strength.

The transition line $B_{OD}(T)$ can also be{\sf \ }derived \cite{25p5} by
considering the energy balance at the transition: The transition occurs when
the sum of pinning energy and thermal energy exceeds the elastic energy
barrier:

$E_{el}=E_{pin}+kT$ \qquad \qquad \qquad \qquad \qquad (2) \newline
where $E_{el}=\varepsilon \epsilon _{o}c_{L}^{2}a_{o}$ is the elastic energy
near the transition line, $E_{pin}=U_{dp}\left( {L_{o}/L_{c}}\right) ^{1/5}$
is the pinning energy of a single vortex\cite{9,10}, and $U_{dp}=(\gamma
\varepsilon ^{2}\epsilon _{o}\xi ^{4})^{1/3}$ is the single vortex depinning
energy. Both approaches Eq. (1) and (2), yield the {\it same} expression for 
$B_{OD}(T)$.

The solution of either Eq. (1) or (2) yields transition lines of different
temperature dependence, depending on the pinning parameter $\gamma $ and on
the anisotropy $\varepsilon $. To demonstrate this variety of behaviors, we
present numerical solutions for $B_{OD}\left( {T}\right) $, fixing $%
\varepsilon $ so that $16\pi ^{2}\lambda _{o}^{2}k/\phi
_{o}^{5/2}\varepsilon c_{L}^{2}=1$ \cite{note} and varying $\Gamma
_{o}=\left( 2\xi _{o}^{6}\varepsilon ^{2}/c_{L}^{2}k^{4}\right) ^{1/5}\gamma
_{o}^{2/5}$, i.e. controlling the pinning strength $\gamma _{o}$. In the
calculations we use the explicit temperature dependences of the coherence
length $\xi =\xi _{o}\left( {1-\left( {T/T_{c}}\right) ^{4}}\right) ^{-1/2}$%
and the penetration depth $\lambda =\lambda _{o}\left( {1-\left( {T/T_{c}}%
\right) ^{4}}\right) ^{-1/2}$. We also consider two pinning mechanisms:
Either ``$\delta T_{c}$ pinning'', caused by spatial fluctuations of the
transition temperature T$_{c}$, or ``$\delta l$ pinning'', caused by
fluctuations of the charge carrier mean free path near a lattice defect\cite%
{25}. In the former case the pinning parameter is $\gamma =\gamma
_{o}^{T}\left( {1-\left( {T/T_{c}}\right) ^{4}}\right) ^{2}$ and in the
latter $\gamma =\gamma _{o}^{l}\left( {1-\left( {T/T_{c}}\right) ^{4}}%
\right) ^{4}$ (Ref. \cite{4}), where either $\gamma _{o}^{T}$ or $\gamma
_{o}^{l}$ replace $\gamma _{o}$ in the expression for $\Gamma _{o}$.

Fig. 2 shows the calculated order-disorder transition line $B_{OD}\left( {T}%
\right) $ (solid curve in the figure), and the irreversibility line $%
B_{irr}(T)$ (dashed curve) estimated by $E_{pin}=kT$, for three different
values of $\Gamma _{o}$, assuming $\delta T_{c}$-pinning mechanism. For
comparison we also show in Fig. 2 the `pure' solid-solid transition line $%
B_{ss}(T)$ (dash-dotted) and `pure' melting line $B_{m}(T)$ (dotted). $%
B_{ss}(T)$ is derived from $E_{el}=E_{pin}$, which neglects the thermal
energy, therefore it is independent of temperature in intermediate
temperature range and descends towards $T_{c}$ as a result of the
temperature dependences of the superconducting parameters\cite{6}. $B_{m}(T)$
is a solution to $E_{el}=kT$, which neglects the pinning energy, therefore
it is unaffected by changes in pinning strength. We maintain that the
experimentally measured transition line - identified by either a jump in
reversible magnetization or the appearance of a second peak in the
irreversible magnetization - corresponds to the $B_{OD}(T)$ curve. Since the
order-disorder transition is driven by {\it both} pinning and thermal
fluctuations, $B_{OD}(T)$ will lie below both $B_{m}(T)$ and $B_{ss}\left( {T%
}\right) $, both of which utilize only one mechanism for the destruction of
the quasi-ordered vortex lattice. The crossing point between $B_{irr}(T)$
and $B_{OD}(T)$ is the ``critical point'' dividing the $B_{OD}(T)$ line into
two segments: The one lying above the irreversibility line will be
manifested by a jump in the reversible magnetization and identified
experimentally as a melting line; the other segment lying below the
irreversibility line will be evinced as a second magnetization peak and
identified experimentally as a solid-solid transition line.

For $\Gamma _{o}=1$ (i.e. relatively small pinning parameter), the effect of
pinning on the order-disorder transition is minor, therefore $B_{OD}(T)$
lies very close to the `pure' melting line $B_{m}(T)$ and retains its
concave shape (Fig. 2{\it a}). $B_{OD}(T)$ crosses the irreversibility line
at extremely low temperatures, so that throughout most of the temperature
range the transition will be manifested as a jump in the reversible
magnetization, as measured in high purity YBCO \cite{26}.

For $\Gamma _{o}=10^{6}$ (relatively large pinning) the effect of
temperature is small, therefore the order-disorder transition line lies near
the `pure' solid-solid transition line $B_{ss}({T)}$ and adopts its convex
shape (see Fig. 2{\it c}), as observed in NCCO\cite{6} (see Fig. 1). In this
case, the intersection of $B_{OD}\left( {T}\right) $ with the
irreversibility line is close to T$_{c}$, so that throughout most of the
temperature range the transition will be evinced as a second magnetization
peak.

For $\Gamma _{o}=500$ (intermediate pinning strength) the deviation of $%
B_{OD}(T)$ from both $B_{m}(T)$ and $B_{ss}\left( {T}\right) $ is marked
(see Fig. 2{\it b}). The shape of the order-disorder transition line $%
B_{OD}\left( {T}\right) $ retains the concave shape of $B_{m}(T)$ but since
most of the transition line lies below the irreversibility line, the
transition will be manifested as a second magnetization peak. This kind of
behavior of $B_{OD}\left( {T}\right) $ was observed in LaSCO\cite{8} (see
Fig. 1).

A non-monotonous behavior can be obtained by invoking $\delta l$-pinning
mechanism, as depicted in Figure 3. In this case, $B_{ss}(T)$ is independent
of temperature at intermediate temperatures, {\it increases} with
temperature and diverges near T$_{c}$. For $\Gamma _{o}=10^{7}$ (relatively
large pinning), the incorporation of thermal fluctuations curbs this ascent,
and results in a peak in $B_{OD}\left( {T}\right) $ as depicted in Fig. 3%
{\it a}. This peak may signify an inverse-melting effect\cite{18,27} as
observed experimentally in BSCCO\cite{19}, YBCO\cite{4,27}, and Pb-BSCCO\cite%
{7}. An alternative explanation \cite{4,11} to the peak in the transition
line attributes this phenomenon to the depinning of the vortices by strong
thermal fluctuations, which smear the pinning potential above the depinning
temperature, $T_{dp}$. This effect was introduced into the expression for
the solid-solid transition through an exponential increase of the Larkin
length above the depinning temperature \cite{4,11}. Our analysis predicts a
peak in $B_{OD}\left( {T}\right) $\ irrespective of the value of $T_{dp}$.

For $\Gamma _{o}=10^{5}$ (lower pinning strength, see Fig. 3{\it b}), two
phenomena are observed: The inverse-melting peak is depressed, and the
critical point moves to lower temperatures. This explains the data of
Khaykovich {\it et al.}\cite{19} and Nishizaki {\it et al}.\cite{5}, showing
that by repeatedly irradiating a crystal the peak in the transition line is
enhanced, and the critical point shifts systematically to higher
temperatures. Furthermore, a dip in the order-disorder transition line
becomes noticeable at intermediate temperatures. Such a dip was previously
reported for YBCO\cite{13,17} (see Fig. 1) and for Pb-BSCCO\cite{7}, and was
attributed to Bean-Livingston barriers\cite{28} or to masking of the
fishtail onset by the field of full penetration\cite{7}. Our analysis shows
that this dip is due to the combined effect of thermally- and
disorder-induced fluctuations in materials where $\delta l$-pinning is the
dominant pinning mechanism. At low temperatures the elastic and pinning
energies are virtually temperature independent, and $B_{ss}\left( {T}\right) 
$ is flat. Thermal fluctuations, however, become stronger as the temperature
is increased causing a deviation of $B_{OD}\left( {T}\right) $ from $%
B_{ss}\left( {T}\right) $. The two lines (Fig. 3{\it b}) merge at low
temperature, but as the temperature is increased, thermal fluctuations allow
the vortices to displace and adjust to the pinning landscape and thereby
induce the order-disorder transition at lower fields. This effect competes
with the thermal dependence of the pinning energy, which stems from the
temperature dependence of the superconducting parameters and causes $%
B_{ss}\left( {T}\right) $ to rise and diverge. At higher temperatures the
latter effect wins, and the transition line $B_{OD}\left( {T}\right) $
increases. Further decrease of the pinning strength to $\Gamma _{o}=1$
results in a monotonously decreasing order-disorder transition line (Fig. 3%
{\it c}).

In conclusion, we have described an order-disorder vortex phase transition
driven by {\it both} thermal fluctuations and disorder-induced fluctuations.
By varying the pinning strength a wide spectrum of transition lines is
obtained, resembling those measured in various high-T$_{c}$ superconductors.
The intersection between the transition line and the irreversibility line
defines a ''critical point`` which divides the transition line into two
segments: One associated with the jump in the reversible magnetization and
identified experimentally as a melting line, and the other associated with
the `fishtail' and identified experimentally as a solid-solid transition
line. For $\delta T_{c}$-pinning, different pinning strengths yield
monotonous transition lines similar to those obtained in clean untwinned
YBCO \cite{26}, LaSCO\cite{8} and NCCO\cite{6}. For $\delta l$-pinning
non-monotonous transition lines are obtained, with a characteristic peak as
observed in YBCO\cite{4}, BSCCO\cite{18,19} and Pb-BSCCO\cite{7}. In
addition, a decrease at low temperature, similar to that observed in YBCO %
\cite{13,17} and Pb-BSCCO\cite{7}, can also be reproduced. The nature of the
disordered phase may be characterized as a liquid, pinned-liquid, or
entangled solid state, depending on the relative contribution of thermal and
disordered induced fluctuations. When thermal (disordered-induced)
fluctuations dominate, the disordered phase exhibits liquid (disordered
solid) characteristics. When both fluctuations are comparable, the
disordered phase behaves as a `pinned-liquid'\cite{8}.

{\it Acknowledgments.} This manuscript is part of Y.R.'s PhD thesis.
Important and stimulating comments from E. Zeldov are acknowledged. Y. R.
acknowledges financial support from Mifal Hapayis - Michael Landau
Foundation. Y. Y. acknowledges support from the US-Israel Binational Science
Foundation. A. S. acknowledges support from the Israel Science Foundation.
This research was supported by The Israel Science Foundation - Center of
Excellence Program, and by the Heinrich Hertz Minerva Center for High
Temperature Superconductivity.

Figure Captions

{\it Figure 1:} Vortex solid-solid transition lines measured in YBCO
(triangles), NCCO (squares) and LaSCO (circles), exhibit different
qualitative behavior. Transition field is normalized to its value at lowest
temperature 20.5 kOe, 260 G, 11.8 kOe respectively; temperature is
normalized by T$_{c}$= 93, 26, 32 K respectively.

{\it Figure 2:} Calculated order-disorder transition line, $B_{OD}\left( {T}%
\right) $ (solid curve), irreversibility line $B_{irr}(T)$ (dashed), `pure'
melting line $B_{m}(T)$ (dotted) and `pure' solid-solid transition line $%
B_{ss}(T)$ (dashed-dotted), for three different values of pinning strength
assuming $\delta T_{c}$-pinning mechanism. Stars mark critical points.

{\it Figure 3:} Same as Fig. 2 assuming $\delta l$-pinning mechanism.


\begin{references}
\bibitem{1} E. Zeldov, D. Majer, M. Konczykowski, V. B. Geshkenbein, V. M.
Vinokur, and H. Shtrikman, Nature {\bf 375}, 373 (1995).

\bibitem{2} B. Khaykovich, E. Zeldov, D. Majer, T. W. Li, P. H. Kes, and M.
Konczykowski, Phys. Rev. Lett. {\bf 76}, 2555 (1996).

\bibitem{3} K. Deligiannis, P. A. J. deGroot, M. Oussena, S. Pinfold, R.
Langan, R. Gagnon, and L. Taillefer, Phys. Rev. Lett. {\bf 79}, 2121 (1997).

\bibitem{4} D. Giller, A. Shaulov, Y. Yeshurun, and J. Giapintzakis, Phys.
Rev. B {\bf 60}, 106 (1999).

\bibitem{5} T. Nishizaki, T. Naito, S. Okayasu, A. Iwase, and N. Kobayashi,
Phys. Rev. B {\bf 61}, 3649 (2000).

\bibitem{6} D. Giller, A. Shaulov, R. Prozorov, Y. Abulafia, Y. Wolfus, L.
Burlachkov, Y. Yeshurun, E. Zeldov, V. M. Vinokur, J. L. Peng, and R. L.
Greene, Phys. Rev. Lett. {\bf 79}, 2542 (1997).

\bibitem{7} M. Baziljevich, D. Giller, M. McElfresh, Y. Abulafia, Y.
Radzyner, J. Schneck, T. H. Johansen, and Y. Yeshurun, Phys. Rev. B {\bf 62}%
, 4058 (2000).

\bibitem{8} Y. Radzyner, A. Shaulov, Y. Yeshurun, I. Felner, J. Shimoyama,
and K. Kishio, Phys. Rev. B (in print).

\bibitem{9} D. Ertas and D. R. Nelson, Physica C {\bf 272}, 79 (1996).

\bibitem{10} V. Vinokur, B. Khaykovich, E. Zeldov, M. Konczykowski, R. A.
Doyle, and P. H. Kes, Physica C {\bf 295}, 209 (1998).

\bibitem{11} T. Giamarchi and P. LeDoussal, Phys. Rev. B {\bf 55}, 6577
(1997).

\bibitem{12} S. B. Roy and P. Chaddah, Physica C {\bf 279}, 70 (1997).

\bibitem{13} Y. Radzyner, S. B. Roy, D. Giller, Y. Wolfus, A. Shaulov, P.
Chaddah, and Y. Yeshurun, Phys. Rev. B {\bf 61}, 14362 (2000)..

\bibitem{14} Y. Yamaguchi, G. Rajaram, N. Shirakawa, A. Mumtaz, H. Obara, T.
Nakagawa, and H. Bando, Phys. Rev. B {\bf 63}, 014504 (2001).

\bibitem{15} Y. P. Sun, Y. Y. Hsu, B. N. Lin, H. M. Luo, and H. C. Ku, Phys.
Rev. B {\bf 61}, 11301 (2000).

\bibitem{16} M. Pissas, E. Moraitakis, G. Kallias, and A. Bondarenko, Phys.
Rev. B {\bf 62}, 1446 (2000).

\bibitem{17} H. Kupfer, T. Wolf, R. Meier-Hirmer, and A. A. Zhukov, Physica
C {\bf 332}, 80 (2000).

\bibitem{18} N. Avraham, B. Khaykovich, Y. Myasoedov, M. Rappaport, H.
Shtrikman, D. E. Feldman, T. Tamegai, P. H. Kes, M. Li, M. Konczykowski, K.
van der Beek, and E. Zeldov, Nature {\bf 411}, 451 (2001).

\bibitem{19} B. Khaykovich, M. Konczykowski, E. Zeldov, R. A. Doyle, D.
Majer, P. H. Kes, and T. W. Li, Phys. Rev. B {\bf 56}, R517 (1997).

\bibitem{20} H. Safar, P. L. Gammel, D. A. Huse, D. J. Bishop, W. C. Lee, J.
Giapintzakis, and D. M. Ginsberg, Phys. Rev. Lett. {\bf 70}, 3800 (1993).

\bibitem{21} C. J. van der Beek, S. Colson, M. V. Indenbom, and M.
Konczykowski, Phys. Rev. Lett. {\bf 84}, 4196 (2000).

\bibitem{23} Y. Y. Goldschmidt, Phys. Rev. B {\bf 56}, 2800 (1997).

\bibitem{23p2} V. Vinokur, B. Khaykovich, and E. Zeldov, unpublished and
private communication.

\bibitem{23p5} J. Kierfeld and V. Vinokur, Phys. Rev. B {\bf 61}, R14928
(2000).

\bibitem{24} G. P. Mikitik and E. H. Brandt, Phys. Rev. B {\bf 64}, 184514
(2001) .

\bibitem{22} A. Soibel, E. Zeldov, M. Rappaport, Y. Myasoedov, T. Tamegai,
S. Ooi, M. Konczykowski, and V. B. Geshkenbein, Nature {\bf 406}, 282 (2000).

\bibitem{25} G. Blatter, M. V. Feigelman, V. B. Geshkenbein, A. I. Larkin,
and V. M. Vinokur, Rev. Mod. Phys. {\bf 66}, 1125 (1994).

\bibitem{note} Increasing or decreasing the anisotropy results in the same
variety of lines depicted in Fig. 2, but which correspond now to different
values of the pinning strength.

\bibitem{25p5} Y. Radzyner {\it et. al}, preprint.

\bibitem{26} A. Junod, M. Roulin, J. Y. Genoud, B. Revaz, A. Erb, and E.
Walker, Physica C {\bf 275}, 245 (1997).

\bibitem{27} S. B. Roy, Y. Radzyner, D. Giller, Y. Wolfus, A. Shaulov, P.
Chaddah, and Y. Yeshurun, cond-mat/0107100 (2001).

\bibitem{28} M. C. de Andrade, N. R. Dilley, F. Ruess, and M. B. Maple,
Phys. Rev. B {\bf 57}, R708 (1998).
\end{references}
\end{document}